\documentclass[fleqn,a4paper,11pt]{article}
\usepackage{color}

\usepackage{graphicx}
\usepackage{amssymb}
\newcommand \mumu {$\mu^+\mu^-$ }

\newcommand \pt {p$_T$ }

\newcommand \sqn {$\sqrt{s_{_{NN}}}$ }

\newcommand \ccbar {$c\overline{c} $ }

\newcommand \raa {R$_{AA}$ }
\newcommand \jpsi {J/$\psi$ }

\def\JPG{{J. Phys}~{\bf G}}

\def\NPA{{Nucl. Phys.}~{\bf A}}

\def\PLB{{Phys. Lett.}~{\bf B}}

\def\PRL{Phys. Rev. Lett.\ }
\def\PR{Phys. Rev.\ }

\def\PRC{{Phys. Rev.}~{\bf C}}

\begin{document}

\title{\bf{From RHIC to LHC: First Lessons}}


\author{Itzhak Tserruya \\
            Weizmann Institute of Science, Rehovot 76100, Israel}

\maketitle

\begin{abstract}
 The first heavy-ion run at the LHC with Pb+Pb collisions at \sqn = 2.76 TeV took place in the fall of 2010. 
 In a short and relatively low luminosity run, the three detectors, ALICE, ATLAS and CMS showcased an impressive performance and produced a wealth of a high quality results. 
This article compares the new LHC results with those accumulated over the last decade at  RHIC, focussing on the quantitative and qualitative
 differences between the different energy regimes of these two facilities.  
 
\end{abstract}


{\centering \section*{1. INTRODUCTION}}
\label{sec:Introduction}
 The first heavy-ion run at the LHC took place in the fall of 2010 with Pb+Pb collisions at \sqn = 2.76 TeV, a factor of $\sim$14 higher than the top RHIC energy of \sqn =~200~GeV,  opening a new energy frontier in the investigation of the strongly interacting quark gluon plasma.  Although the run was relatively short with an integrated luminosity of $\sim$9~$ֿ\mu$b$^{-1}$ the three experiments, ALICE, ATLAS and CMS, reported shortly after the run an impressive amount of high quality results \cite{qm11-proc}.
 
 The prime question being asked with the advent
 of LHC results is what is different at the LHC compared to the knowledge accumulated over the last decade at RHIC? There are obvious quantitative differences between the two cases but the interesting issue is whether there are qualitative differences. In these proceedings, I shall attempt a comparative discussion of results from RHIC and LHC. The discussion is restricted to a personal selection of the available results as it is clearly not possible to cover them all in the frame of this short paper.
\vspace{5mm}

{\centering \section*{2. GLOBAL OBSERVABLES}}
\label{sec:global}
 
\begin{figure}[h!]
     \includegraphics[width=72mm]{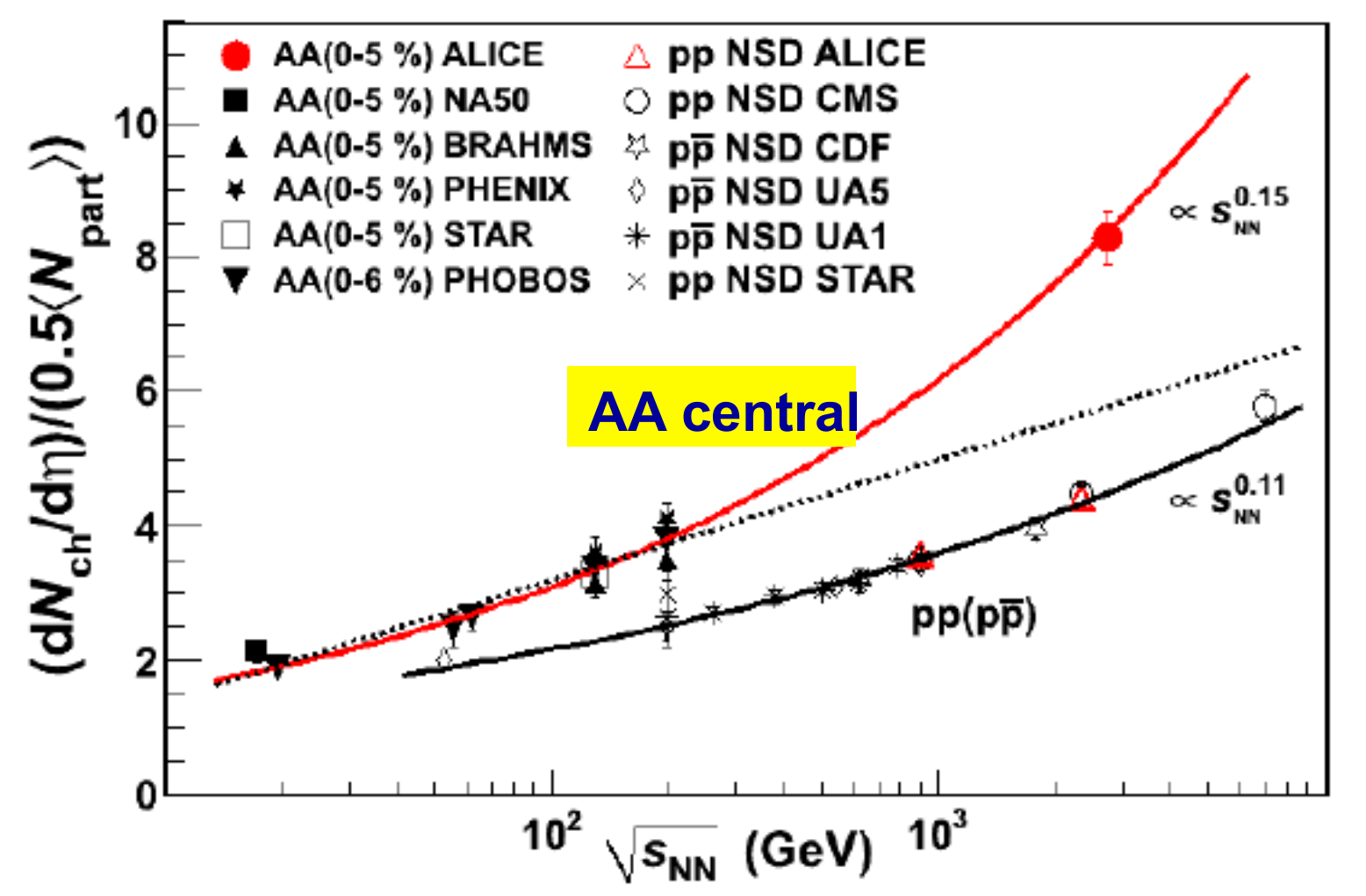}
    \includegraphics[width=72mm]{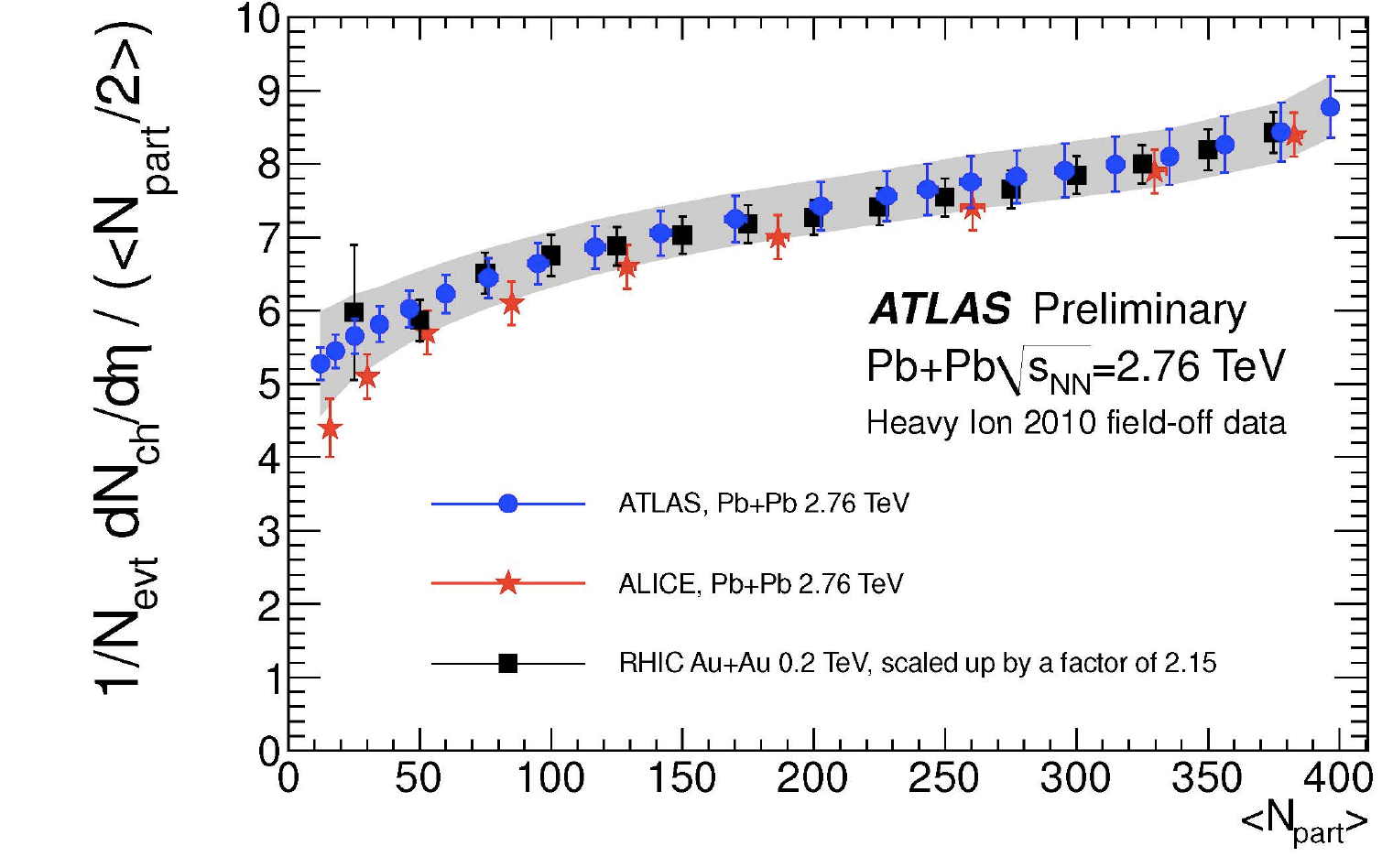}
    \caption{Left panel: Energy dependence of the charged particle density at mid-rapidity per pair of participants in pp and central AA collisions \cite{alice-multiplicity}. Right panel: Centrality dependence of the charged particle density per pair of participants in AA collisions at LHC and RHIC \cite{steinberg-qm11}. The latter has been scaled up by a factor of 2.15 for a better shape comparison with the LHC data.}
  \label{fig:dndeta}
 \end{figure}

The charged particle density at mid-rapidity per participant pair, in central AA and pp collisions is shown in the left panel of Fig. \ref{fig:dndeta} \cite{alice-multiplicity}. 
 The empirical logrithmic scaling (dotted line) that appeared to work so well at lower energies, up to the top RHIC energy, is clearly ruled out by the new LHC data point.
 The particle density per participant pair, dN$_{ch}$/d$\eta$/0.5N$_{part}$, at 2.76 TeV is a factor of $\sim$2.15 higher than at RHIC and provides a new constraint on models that attempt to describe nuclear collisions.
The transverse energy density, dE$_T$/d$\eta$/0.5N$_{part}$, is found to be a factor of 2.7 higher at LHC than at RHIC, consistent with the increase in particle density and the $\sim$20\% increase in the average particle p$_T$ \cite{toia-qm11}. This translates into an energy density that is estimated to be at least a factor of 3 larger at LHC than at RHIC. The particle density in central AA collisions increases faster than
 in pp collisions. Both are well described by a power law scaling, $\propto s^{0.15}$ and  $\propto s^{0.11}$, in central AA and pp collisions, respectively. 

The centrality dependence of the charged particle density per pair of participants at LHC is very similar to that found at RHIC (see right panel of Fig. \ref{fig:dndeta}) \cite{steinberg-qm11} which itself is very similar to the one found at top SPS energy. The shape similarity thus appears as a robust feature over more than two orders of magnitude in energy 
 casting serious doubts about the validity of the two-component model that attempts to explain particle production as a superposition of a soft component proportional to N$_{part}$ and a hard component proportional to N$_{coll}$; \(dN_{ch}/d\eta = N [f N_{part} + (1-f)N_{coll}] \) should be considered as a successful parametrization of the data. The almost constant value of \(f\) from SPS to LHC could reflect that some saturation scale is at play already at the SPS or that geometrical effects dictate the shape of the $dN_{ch}/d\eta$ centrality dependence.

Measurements of two pion Bose-Einstein correlations (HBT) show an increase in the radii R$_{out}$, R$_{side}$ and R$_{long}$  above the RHIC values that indicate an increase of the freeze-out volume by a factor of about 2 \cite{alice-hbt}.

\vspace{5mm}
{\centering \section*{3. FLOW}}
\label{sec:flow}
\vspace{3mm}

\begin{figure}[h!]
     \includegraphics[width=72mm]{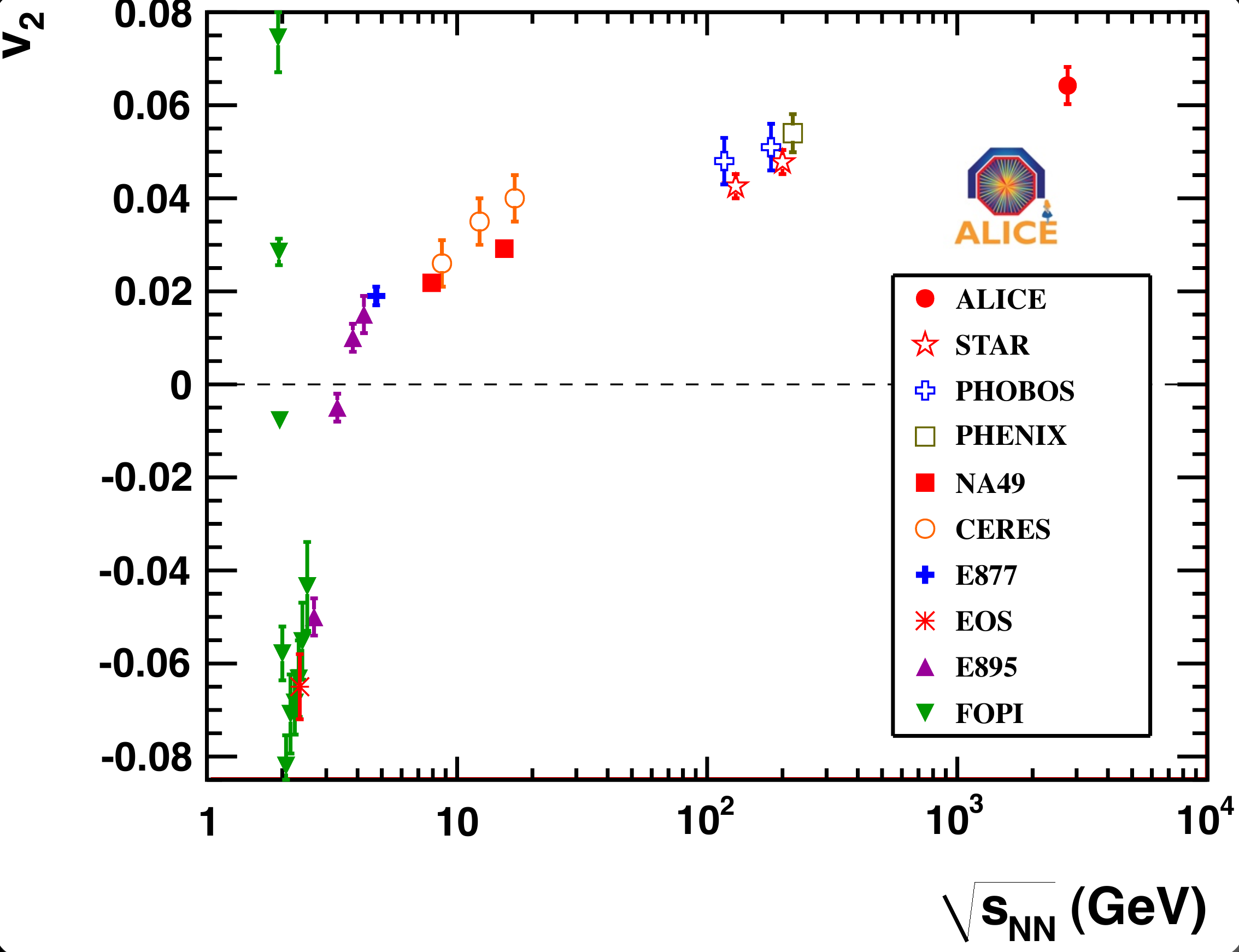}
    \includegraphics[width=65mm]{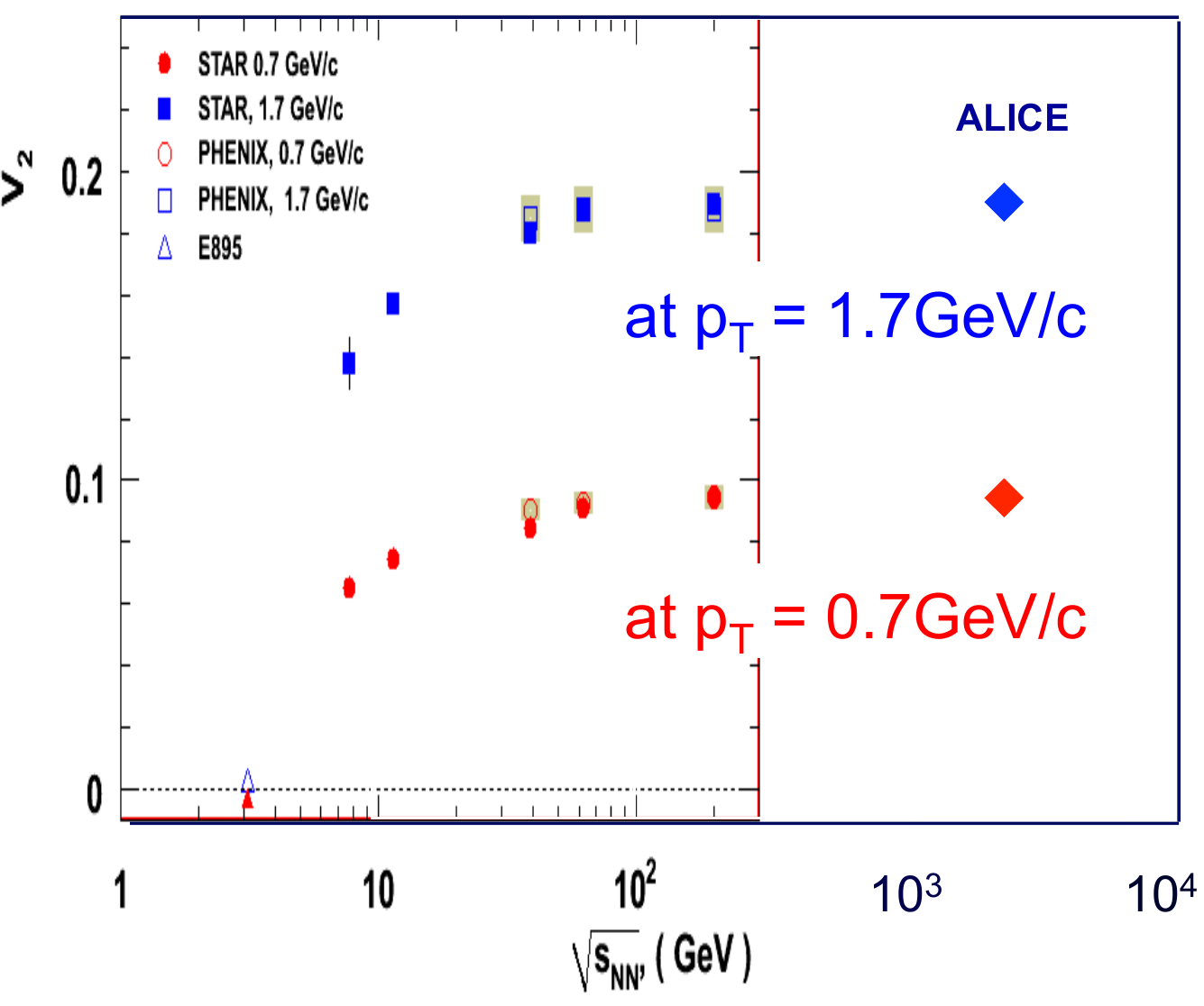}
  \caption{Average (left panel) \cite{alice-flow} and differential (right panel) \cite{shinichi-qm11, gong-qm11} elliptic flow vs $\sqrt{s}$.}
  \label{fig:v2-roots}
 \end{figure}

The average elliptic flow (v$_2  = \int {N(p_T) v_2(p_T) dp_T} / \int {N(p_T) dp_T}$)\footnote{This quantity is often referred to as the integrated $v_2$ but it represents in fact the average value of $v_2$.} in the 20-30\% centrality class increases by  $\sim$25\% from RHIC to LHC (left panel of Fig. \ref{fig:v2-roots}) \cite{alice-flow}. This increase mainly reflects the observed increase in the average p$_T$ rather than an increase of the differential elliptic flow $v_2(p_T)$. The latter 
 changes very little
from $\sqrt{s_{NN}}$~=~39~GeV up to the new LHC data point as illustrated in the right panel of Fig.~\ref{fig:v2-roots} that shows the differential elliptic flow for two p$_T$ bins (0.7 and 1.7 GeV/c)  vs \sqn \cite{shinichi-qm11, gong-qm11}. 
The saturation of $v_2$ at around or below 39 GeV suggests that the perfect liquid property of the QGP discovered at RHIC is valid from this low energy up to at least 2.76 TeV.

 \newpage The valence quark scaling observed at RHIC seems also to work at the LHC although deviations are observed for protons \cite{snellings-qm11}.

\begin{figure}[h!]
     \includegraphics[width=80mm]{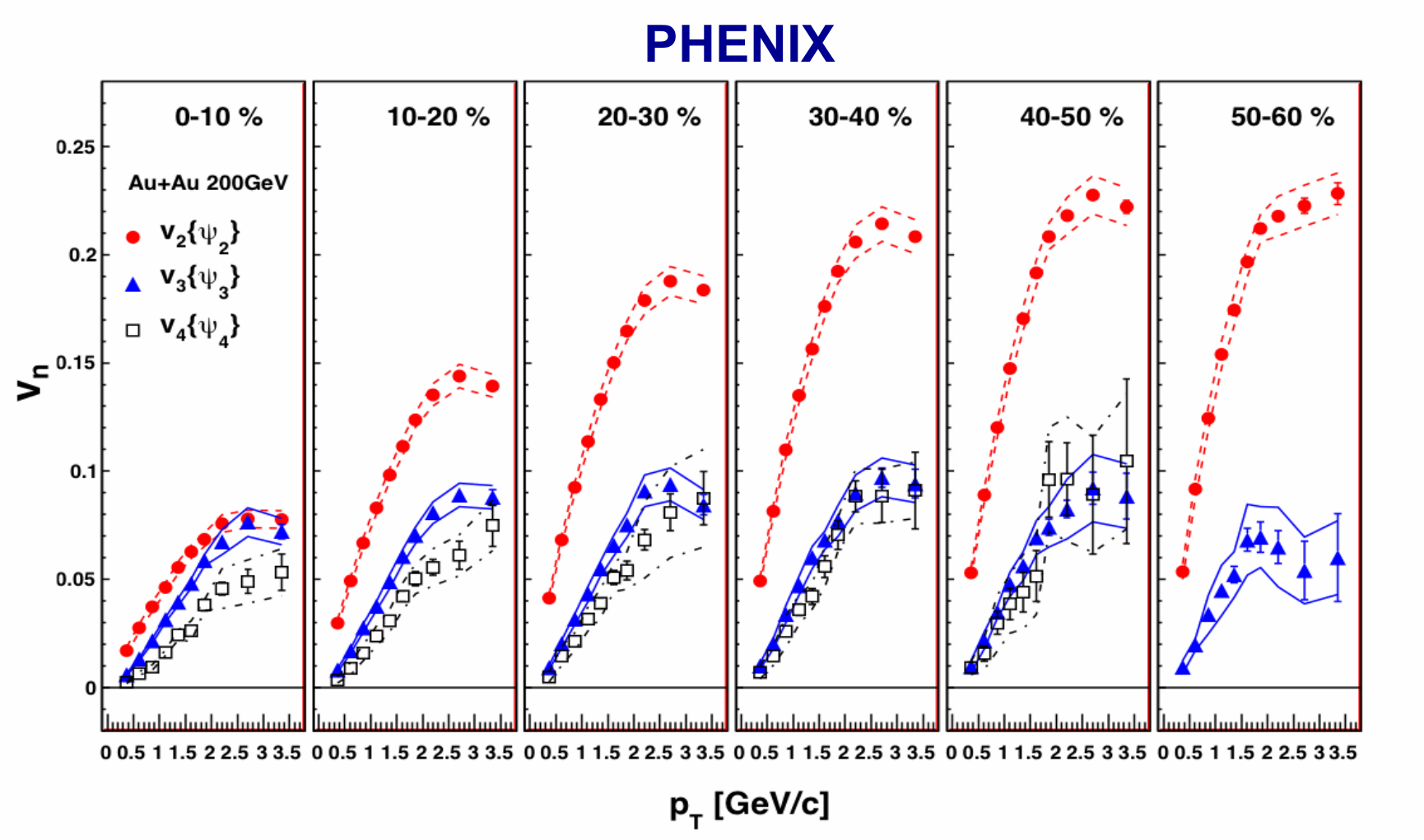}
    \includegraphics[width=72mm]{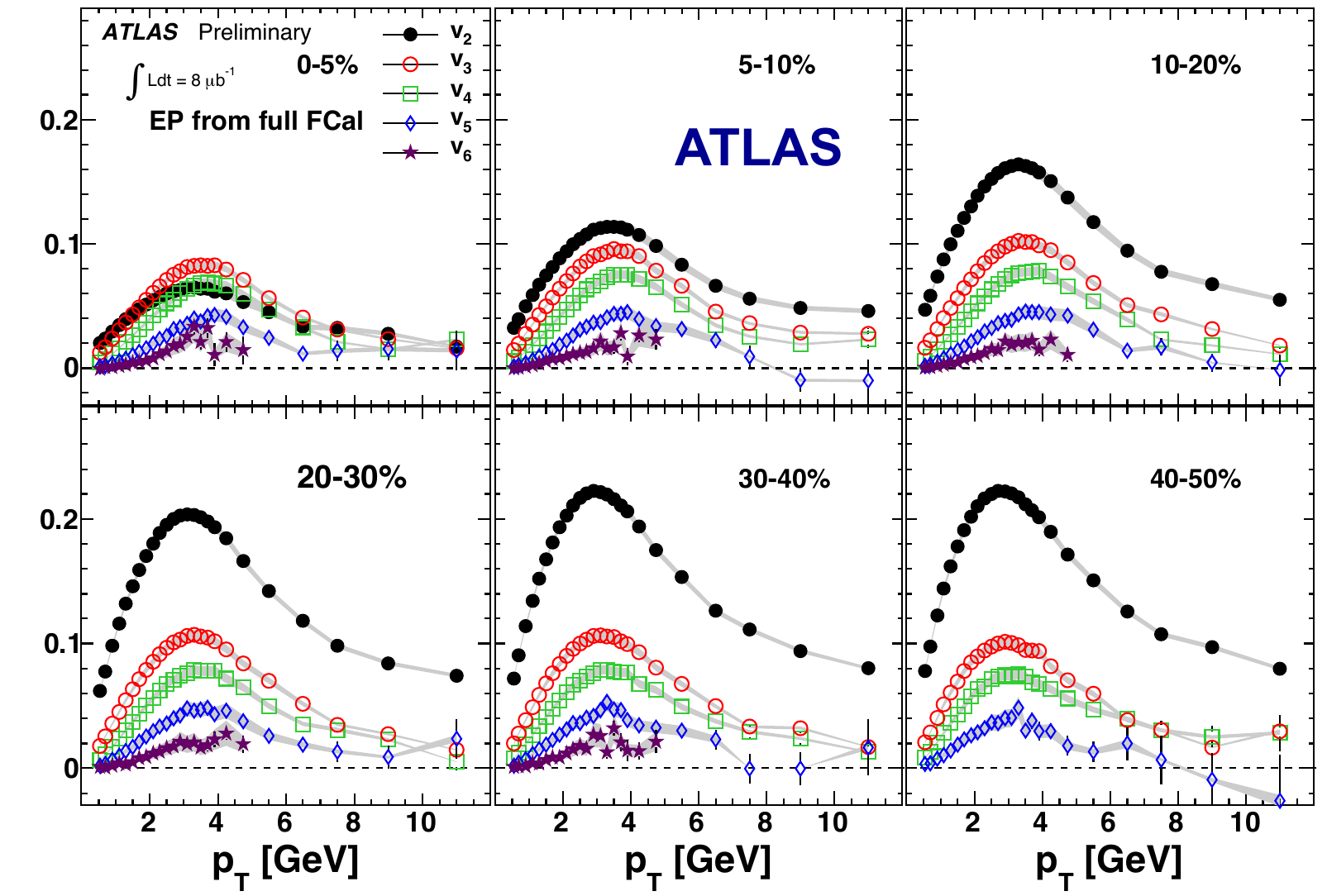}
    \caption{Higher order flow harmonics measured by PHENIX (left panel) \cite{shinichi-qm11} and ATLAS (right panel) \cite{jia-qm11}.}
  \label{fig:high-harmonics}
  \vspace{-6mm}
 \end{figure}

The remarkable lesson on flow from the last Quark Matter conference is the importance of the higher order harmonic components.
Triggered by  recent theoretical work \cite{alver,ma}, the large five experiments (PHENIX and STAR at RHIC and ALICE, ATLAS and CMS at LHC) presented compelling evidence for the importance of the higher order flow components.  
The characteristic features appear very similar at RHIC and LHC (see Fig.~\ref{fig:high-harmonics}) \cite{shinichi-qm11, jia-qm11}: 
sizable $v_n$ up to the sixth order;
same pattern for all n: $v_n$ rises up to $\sim$3~GeV/c and then falls at higher p$_T$;
weaker or no centrality dependence of $v_3$ - $v_6$ as compared to $v_2$ that exhibits a strong centrality dependence.
The saturation of $v_2$  mentioned above seems also to hold for the higher flow components. 

The importance of the higher order harmonics has crucial consequences.  In two particle correlations, the long range $\Delta \eta$ near-side correlations (the so-called ridge) and the double peak structure in the away-side correlations (the shoulder interpreted as Mach cone)  both largely disappear once the higher order flow components are subtracted \cite{shinichi-qm11, jia-qm11}. It is also hoped that the higher order harmonics and in particular $v_3$ will help constraining the viscosity over entropy ratio $\eta$/s. 
  
{\centering \section*{4. PARTICLE SPECTRA AND \raa}}
\label{sec:raa}

\begin{figure}[h!]
     \includegraphics[width=72mm]{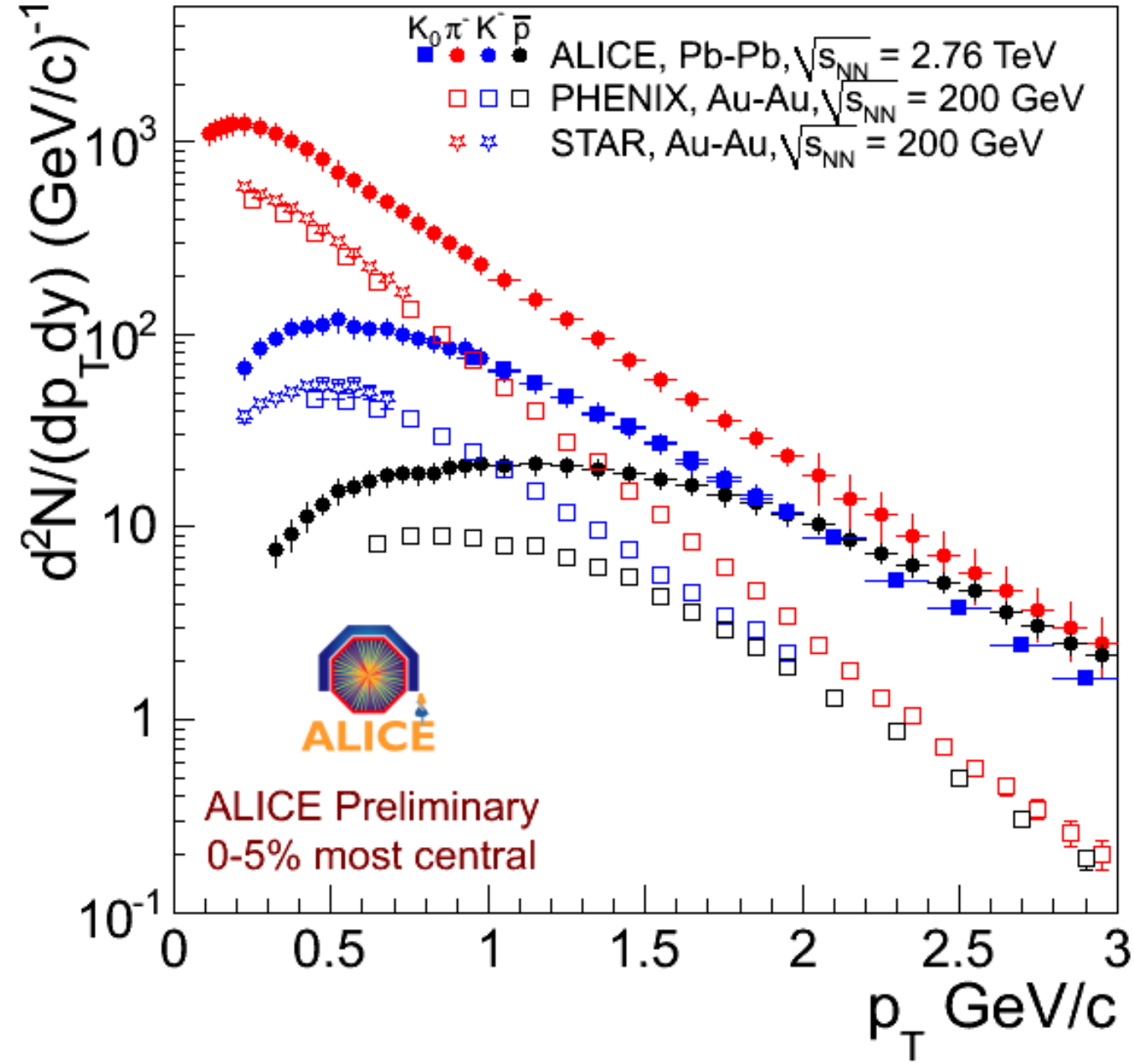}
     \includegraphics[width=72mm, height=68mm]{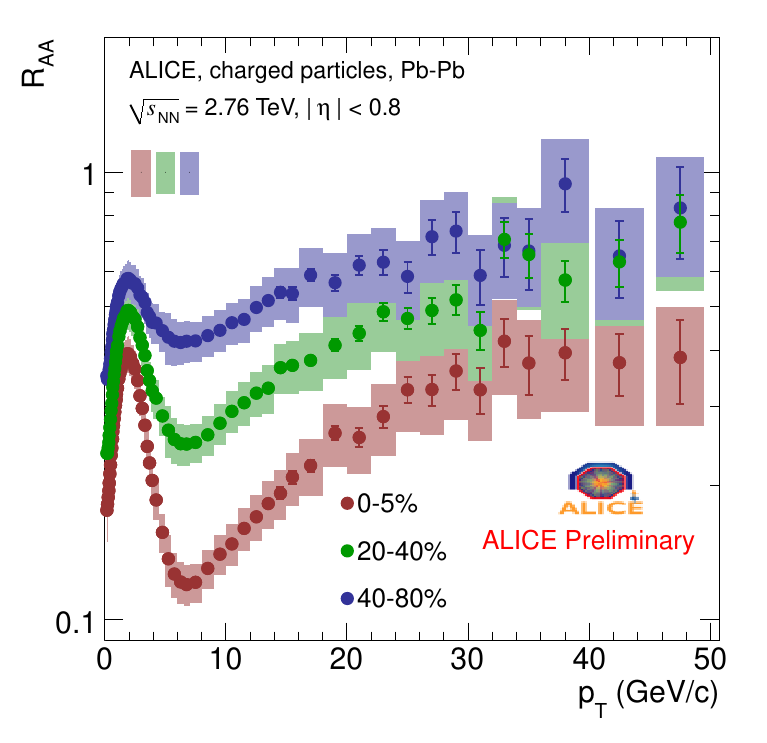}
\caption{Left panel: Identified particle spectra at \sqn = 200 GeV from PHENIX and STAR and at \sqn = 2.76 TeV from ALICE \cite{floris-qm11}. Right panel: \raa of charged particles for three centrality bins measured by ALICE in Pb+Pb collisions at \sqn = 2.76 TeV \cite{appelshauser-qm11}.}
  \label{fig:spectra-raa}
 \end{figure}

Identified particle spectra at RHIC and LHC are compared in the left panel of Fig. \ref{fig:spectra-raa} \cite{floris-qm11}. There is a significant change in slope, the spectra being harder at LHC than at RHIC. There is also a large increase in the particle production cross section at LHC compared to RHIC, which is more prominent at high \pt. The factors are huge. For example, at \pt =3, 10 GeV/c cross sections are larger at LHC than at RHIC by factors of $\sim$10, 50, respectively. Consequently, particle spectra 
 could be measured at the LHC already in the first low-luminosity heavy-ion run with 
 unprecedented \pt reach up to 100 GeV/c.
  
\begin{figure}[h!]
     \includegraphics[width=72mm, height=55mm]{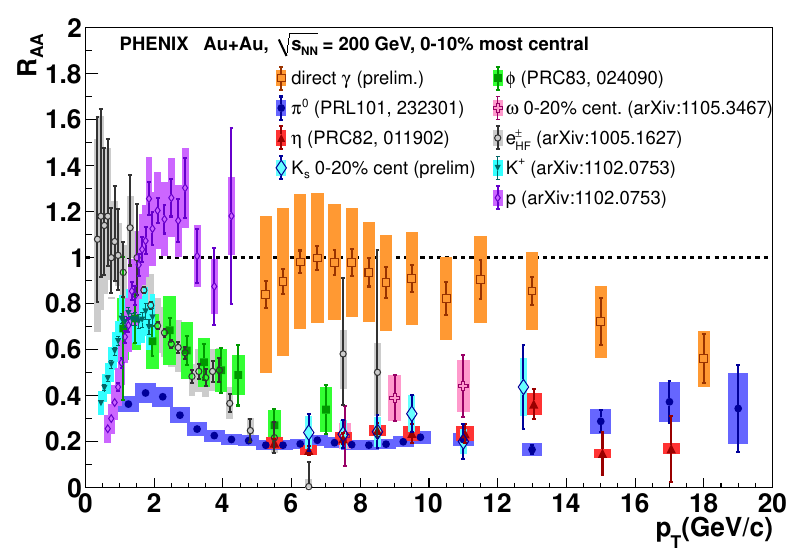}
    \includegraphics[width=72mm]{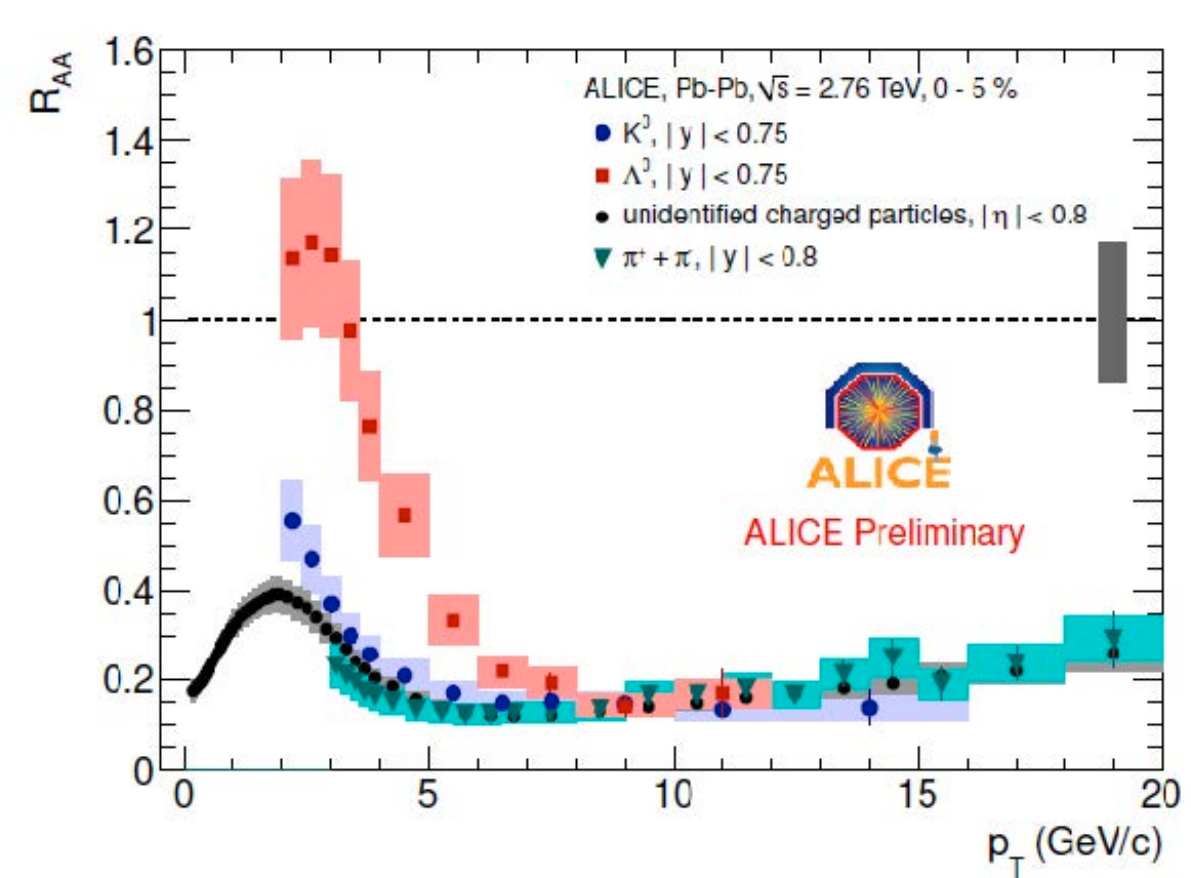}
    \caption{Left panel: \raa for baryons, strange mesons, e from heavy flavor,  light quark mesons and direct photons measured by PHENIX in 0-10\% most central Au+Au collisions at \sqn = 200 GeV  \cite{sharma-qm11}. Right panel: \raa of $\pi$, K, $\Lambda$
 and unidentified charged particles measured by ALICE in central Pb+Pb collisions at \sqn = 2.76 TeV \cite{appelshauser-qm11}.}
  \label{fig:raa-all-particles}
\vspace{5mm}
 \end{figure}

Nuclear modification factors, R$_{AA}$, at LHC are displayed in the right panel of Fig. \ref{fig:spectra-raa}. 
For all centrality bins the ALICE data show the same behavior. At low p$_T$,  \raa exhibits first a fast rise followed by a gradual decrease, reaching  maximum suppression at p$_T$~=~6-7~GeV/c. At \pt \(>\) 7 GeV/c, \raa steadily increases up to $\sim$30 GeV/c  and appears to saturate at higher \pt values.
 In the \pt region of overlap, RHIC data (see e.g. the $\pi^0$ data in the left panel of Fig. \ref{fig:raa-all-particles}) show a similar pattern, although the minimum at \pt~=~6-7~GeV/c is less deep and the rise at \pt\(>\) 7 GeV/c is not clearly established within the experimental uncertainties.
 
The PHENIX results show an interesting hierarchy in the suppression pattern of identified particles at low \pt \(<\) 6-7 GeV/c (left panel of Fig. \ref{fig:raa-all-particles}) \cite{sharma-qm11}.  Light quark mesons show the largest suppression whereas baryons have very small or no suppression at all.  Strange mesons and electrons from heavy flavor ($e_{HF}$) show intermediate suppression:   
\newline 
{\centering \raa (light quark mesons) \(<\) \raa (strange mesons and $e^\pm$ from HF) \(<\) \raa (baryons) }
\newline
On the other hand, at higher p$_T$, all particles, baryons and mesons (independently of their quark flavor) seem to exhibit the same suppression level. The limited amount of ALICE results available so far seem to be consistent with a similar pattern (see right panel of Fig.~\ref{fig:raa-all-particles}) \cite{appelshauser-qm11}. 
 
{\centering \section*{5. JETS AND CORRELATIONS}}
\label{sec:jets}
\vspace{4mm}

There are distinct features between the jet studies at RHIC and LHC. At RHIC, jet studies are practically limited to energies below $E_{jet}$~=~30-50~GeV, the limit being imposed by the cross section. Furthermore, jet information is mainly derived via particle correlations. It is only recently that both PHENIX and STAR were able to obtain results based on full jet reconstruction. 

At LHC, on the other hand, jet measurements are carried out for high energy jets (E$_{jet}$~\(>\)~25~GeV) which are prominently visible and can be fully reconstructed, whereas low energy jets are not due to fluctuations in the underlying event. An interesting question is whether the quench phenomena of low energy jets observed at RHIC are qualitatively different from the high energy jets observed at LHC. In an attempt to answer this question the main characterictic features of jets at RHIC and LHC are compared below.

$\bullet$ \underline {Jet suppression.}
At LHC, the jet yield is suppressed in central collisions by about a factor of 2 and the suppression level is independent of jet energy (see jet R$_{CP}$ measurements in \cite{steinberg-qm11}).
A similar level of suppression is observed at RHIC  (see e.g. the jet \raa measurements in \cite{phenix-jets-raa}).  

\begin{figure}[h!]
     \includegraphics[width=65mm, height=55mm]{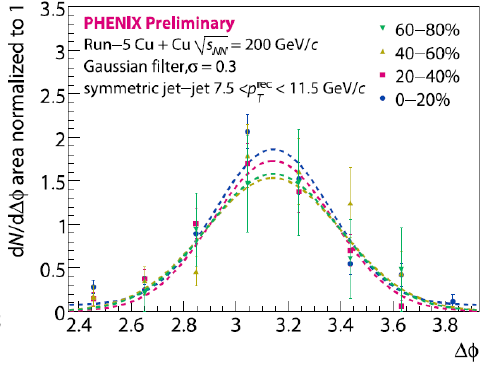}
     \includegraphics[width=85mm, height=55mm]{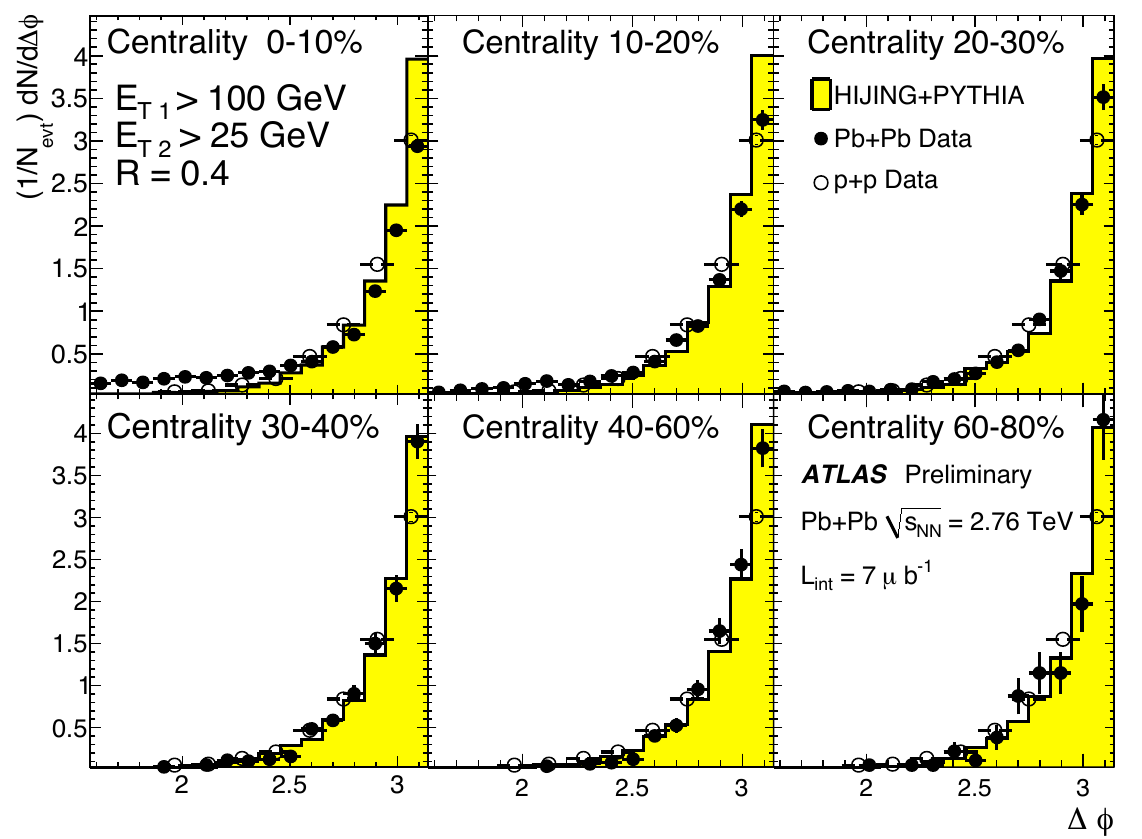}
            \caption{Angular correlations of dijets ($\Delta \phi = \phi_{jet1} - \phi_{jet2}$) in Cu+Cu collisions at RHIC from PHENIX (left panel) \cite{phenix-dijet-correl} and in Pb+Pb collisions at LHC from ATLAS (right panel) \cite{atlas-jets}.}
     \label{fig:jets-back-to-back}
 \end{figure}

$\bullet$ \underline {Dijet correlations.}
Dijets are mostly back-to-back. The same angular correlations as in pp are observed in AA collisions for all centralities. There is no evidence for deflection of the away-side jet, both at the LHC and RHIC as shown in Fig. \ref{fig:jets-back-to-back} \cite{phenix-dijet-correl, atlas-jets}.

\begin{figure}[h!]
\begin{center}
     \includegraphics[width=72mm]{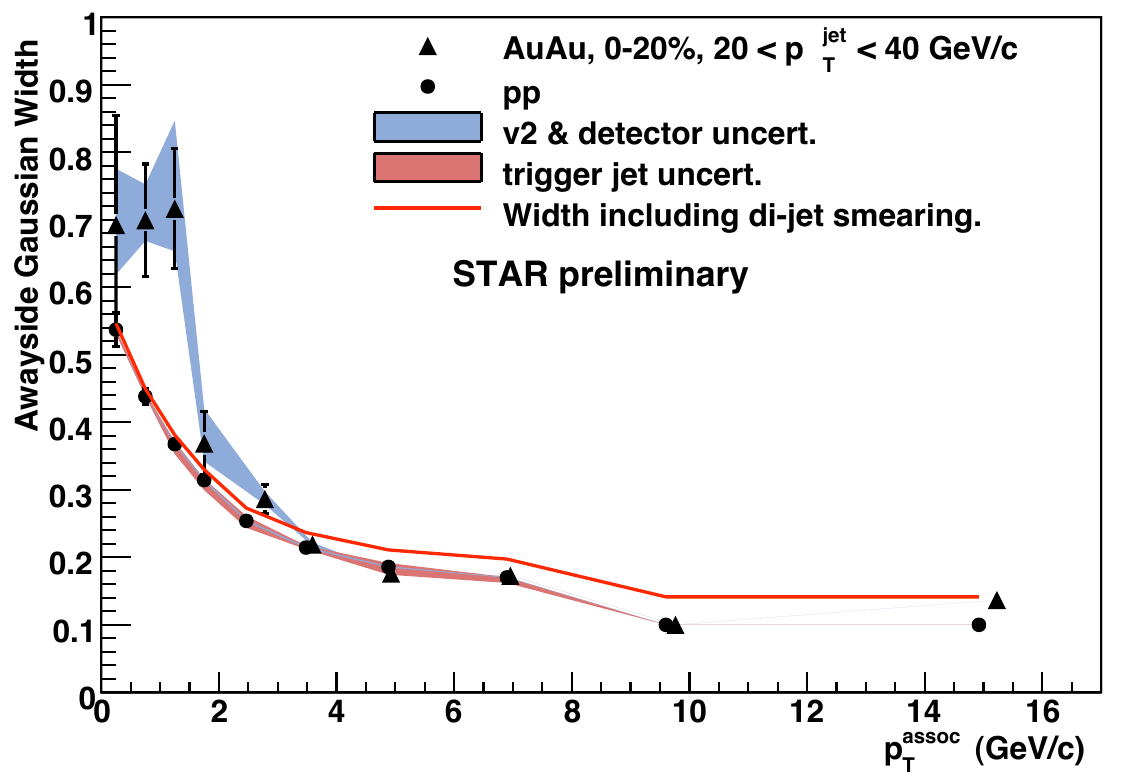}
\end{center}
            \caption{Width of the away-side jet in jet-hadron correlations in central Au+Au collisions as function of \pt of the associated particles \cite{caines-qm11}.}
     \label{fig:jet-hadron-correl-star}
 \end{figure}

$\bullet$ \underline {Jet broadening.}
The away-side jet is broadened at RHIC as illustrated in Fig.~\ref{fig:jet-hadron-correl-star} that shows results derived from jet-hadron correlations measured by STAR in central Au+Au collisions \cite{caines-qm11}. The width of the away-side jet is the same as in pp collisions when selecting high \pt associated particles, whereas it is considerably larger, by almost a factor of 2, when selecting low \pt particles. This suggests that the energy lost by the parton as it traverses the medium appears in soft hadrons that remain correlated with the original parton direction. There is not yet direct evidence of jet broadening at the LHC.

\begin{figure}[h!]
\begin{center}
     \includegraphics[width=75mm]{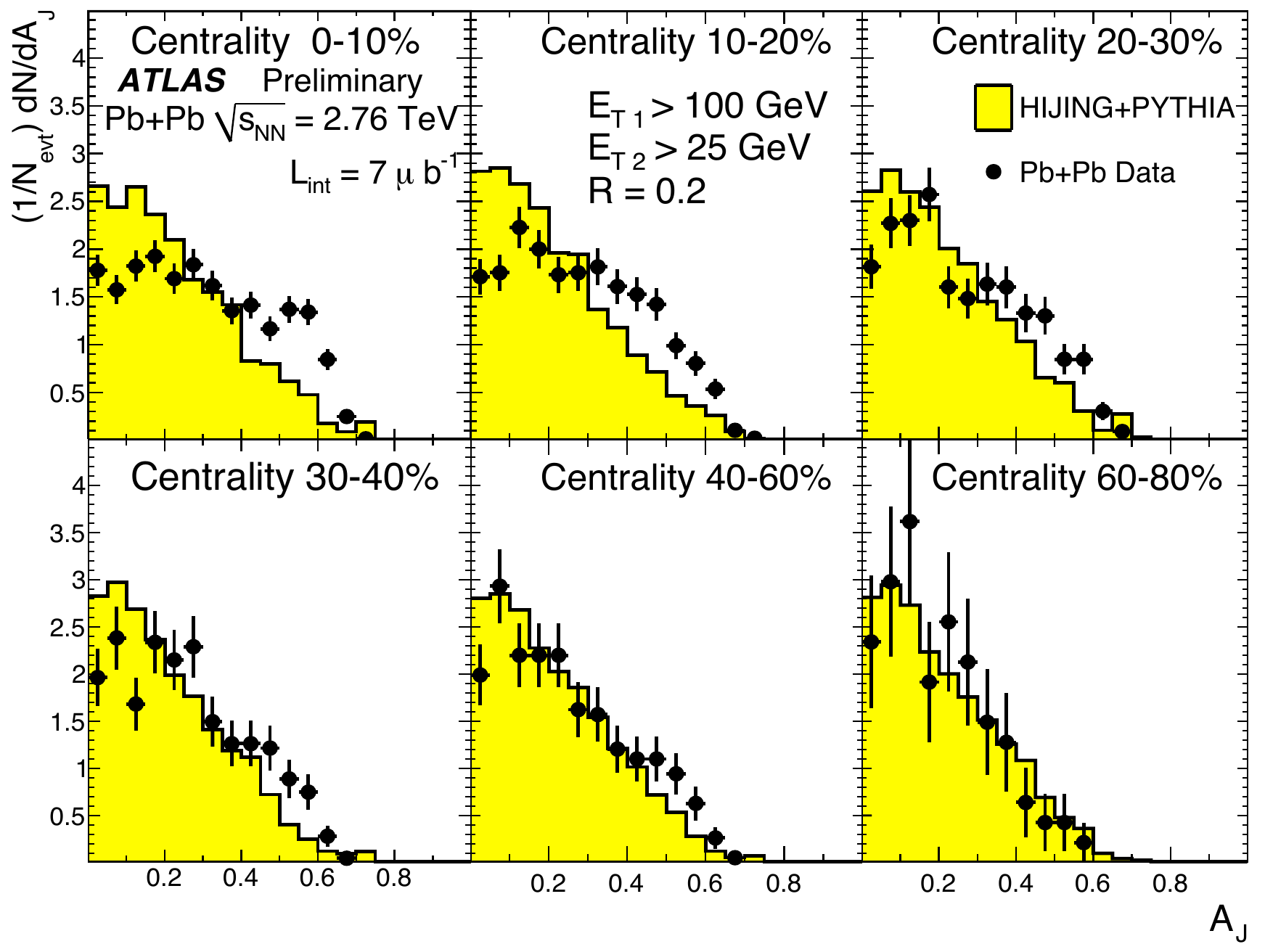}
\end{center}
     \caption{Dijet transverse energy balance characterized by the asymmetry parameter A$_j$=\((E_{T1}-E_{T2}) / (E_{T1}+E_{T2})\) where \(E_{Ti}\) is the transverse energy of jet i. From ATLAS \cite{atlas-jets}. }
                     \label{fig:jets-momentum-balance-atlas}
  \end{figure}

\begin{figure}[h!]
\begin{center}
      \includegraphics[width=140mm]{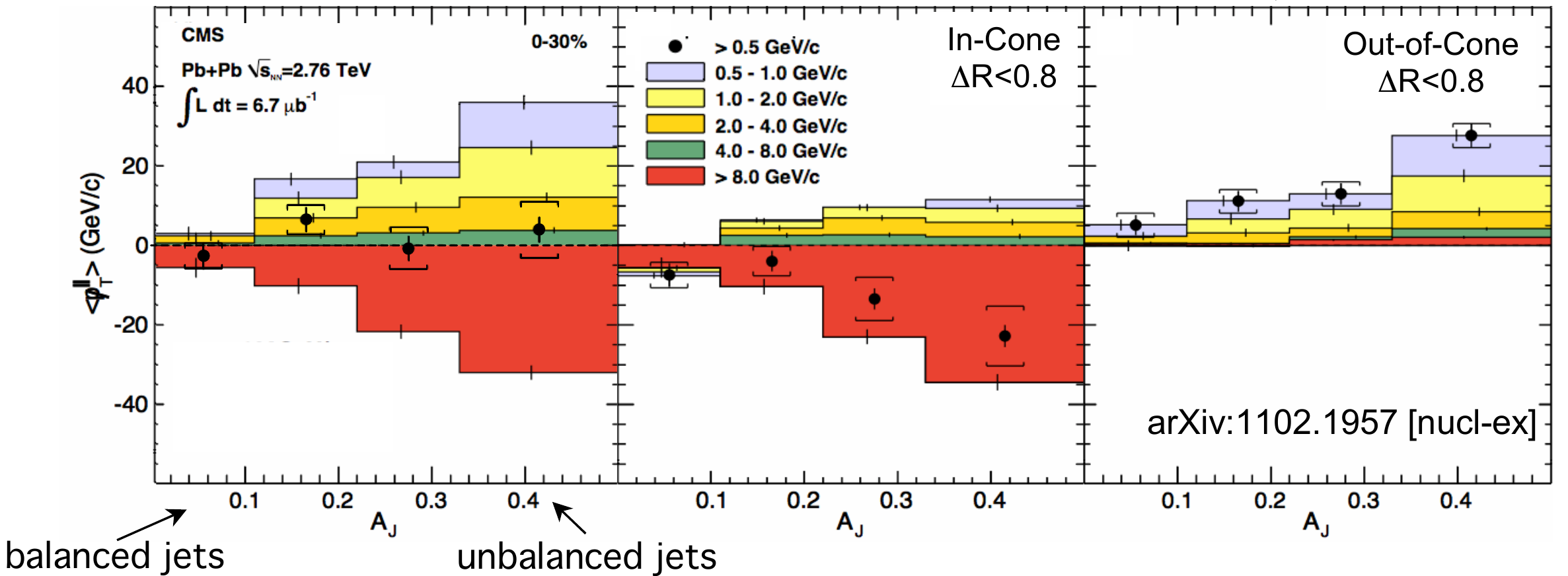} 
\end{center}
   \caption{Dijet momentum balance characterized by the parameter $<p_T^\parallel>$= \(\sum_{track}  -\)
\newline \(p_T^{track} cos(\phi_{track} -\phi_{leading jet})\) vs the asymmetry parameter A$_j$=\( (p_{T1}-p_{T2}) / (p_{T1}+p_{T2}) \). The whole event is momentum balanced (left panel). For tracks inside the jet cones, there is a momentum imbalance that manifests itself as an excess of high-\pt tracks in the leading jet direction (middle panel) and is compensated by an excess of low-\pt tracks outside the cones towards the recoiling jet side (right panel). From CMS \cite{cms-jet-balance}.}
                     \label{fig:jets-where-energy-goes-cms}
 \end{figure}

\newpage
$\bullet$  \underline {Energy and momentum balance.}
At LHC, dijets exhibit a large transverse energy imbalance that grows with centrality as first shown by ATLAS (see Fig. \ref{fig:jets-momentum-balance-atlas}) \cite{atlas-jets}). The imbalance is compensated by low \pt particles (0.5\(<\) \pt \(<\) 2 GeV/c) emitted mostly at large angles with respect to the away-side jet, out of the jet cone, as illustrated by the CMS results shown in Fig. \ref{fig:jets-where-energy-goes-cms} \cite{cms-jet-balance}.

\begin{figure}[h!]
      \includegraphics[width=75mm]{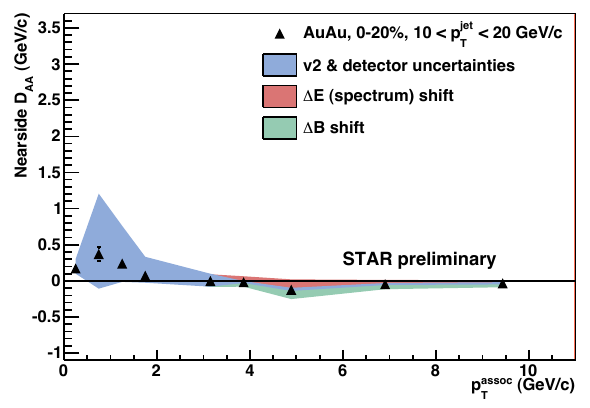}
     \includegraphics[width=75mm]{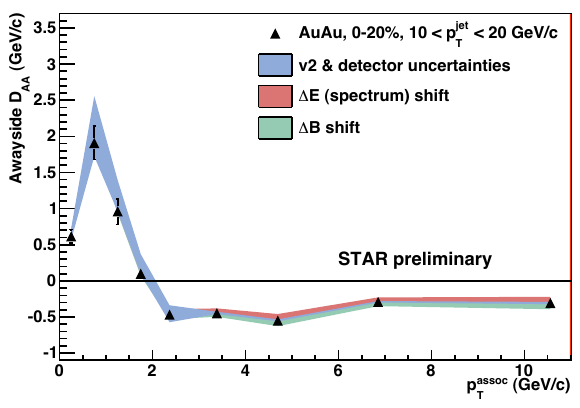}
   \caption{Near-side (left) and away-side (right) $D_{AA}$ distributions as a function of the associated hadron transverse momentum \(p_T^{assoc}\) in central Au+Au collisions measured by STAR \cite{caines-qm11}.}
                     \label{fig:jets-momentum-difference-star}
 \end{figure}

Similar studies are conducted at RHIC \cite{caines-qm11}. In Fig. \ref{fig:jets-momentum-difference-star}, the transverse momentum difference between AA and pp collisions, \( D_{AA}(p_T^{assoc}) = [Y_{AA}(p_T^{assoc}) - Y_{pp}(p_T^{assoc})] p_T^{assoc} \) (where \(Y_{aa}(p_T^{assoc})\) represents the yield of associated tracks with transverse momentum \(p_T^{assoc}\) in aa collisions) is plotted for the near-side (left panel) and the away-side (right panel) jets. Whereas for the near-side jet, the momentum profiles in central Au+Au and pp collisions are very similar, for the away-side jet there is a deficit of high \pt tracks that is within errors compensated by an excess of low \pt tracks \cite{caines-qm11} (the integral of $D_{AA}$ is equal to  \( 1.6_{-0.4-0.4}^{+1.6+0.5}\) GeV/c). One notices here a difference with the LHC results. Whereas at the LHC the momentum balance in the away-side comes to a large extent from low-\pt tracks outside the jet cone, at RHIC good compensation is achieved by the excess of low \pt tracks inside the jet cone.  

\begin{figure}[h!]
 \begin{center}
    \includegraphics[width=90mm, height=50mm]{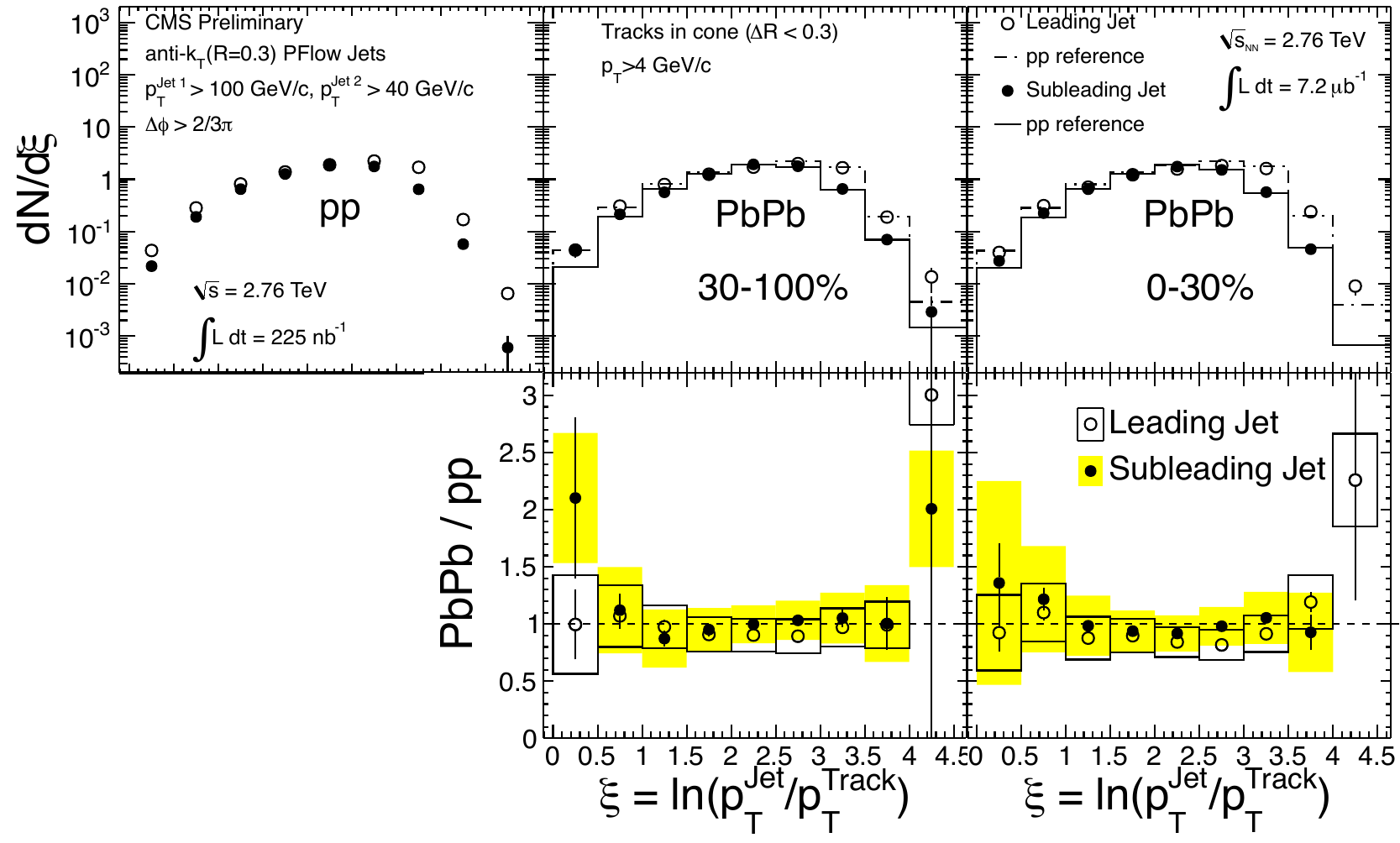}
\end{center}
\vspace{-4mm}
             \caption{Fragmentation functions in pp and for several centrality bins in Pb+Pb collisions measured at LHC by CMS using jet-jet correlations \cite{bolek-qm11}.}
                     \label{fig:jets-ff-cms}
 \end{figure}

\begin{figure}[h!]
 \begin{center}
    \includegraphics[width=72mm, height=50mm]{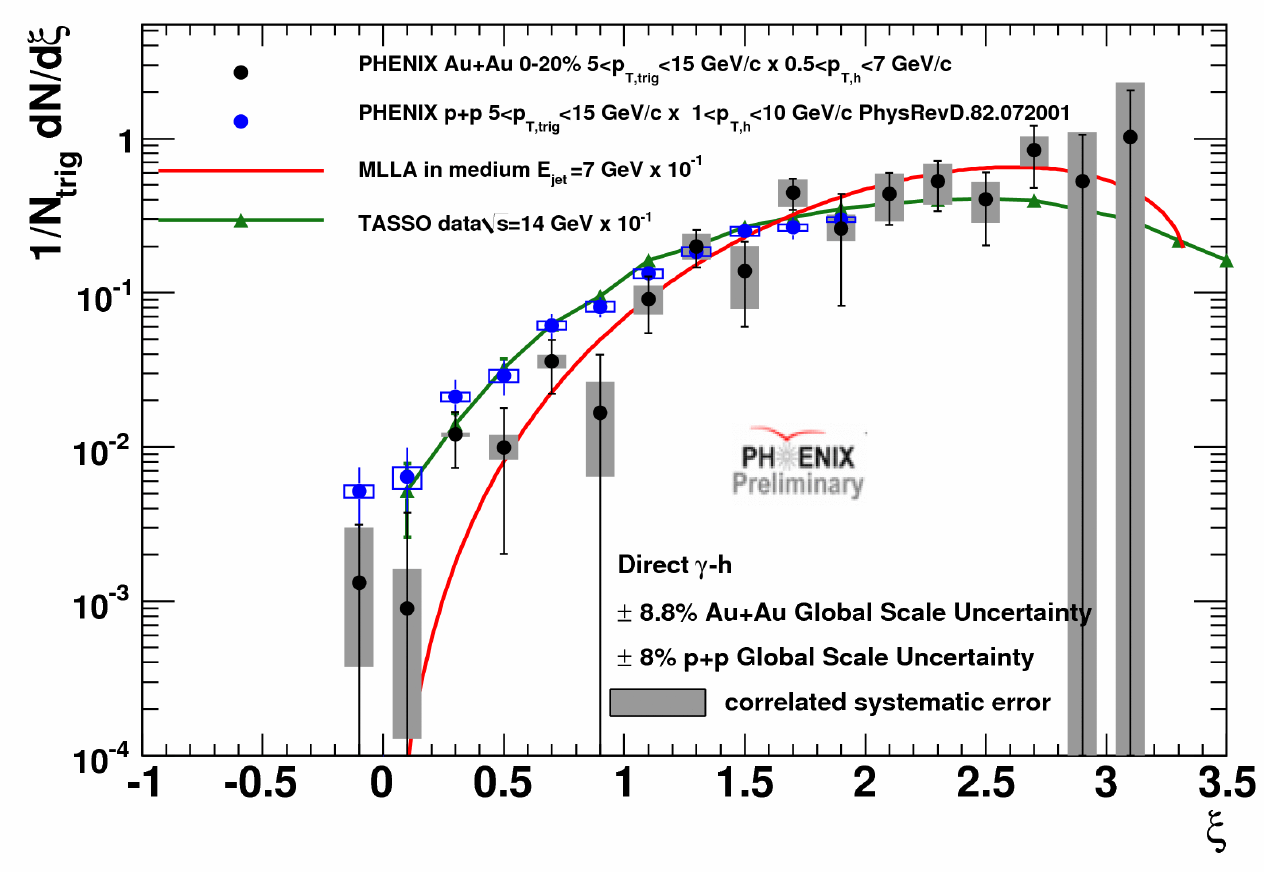}
\end{center}
\vspace{-4mm}
            \caption{Fragmentation functions in pp and central Au+Au collisions measured at RHIC by PHENIX using $\gamma$-jet correlations where the jet energy is determined by the $\gamma$ energy \cite{phenix-jets-raa}.}
                     \label{fig:jets-ff-phenix}
 \end{figure}

\newpage
 $\bullet$  \underline {Fragmentation functions.}
Both ATLAS and CMS report no change in the fragmentation functions measured in Pb+Pb collisions (see e.g. the CMS results in Fig. \ref{fig:jets-ff-cms} \cite{bolek-qm11}). This is to be contrasted with the RHIC results which show that the fragmentation function in Au+Au is modified compared to pp, with a smaller yield at low $\xi$ and a higher yield at high $\xi$ (see Fig.~\ref{fig:jets-ff-phenix} \cite{phenix-jets-raa}). This difference could be due to the different selection cuts (in the results shown in Fig. \ref{fig:jets-ff-cms}, the CMS analysis is restricted to tracks with \pt \(>\) 4 GeV/c~\footnote{ATLAS with a track \pt cut \(>\) 2 GeV/c sees also unmodified fragmentation functions.} whereas the PHENIX analysis uses a much lower \pt cut $>$ 0.5 GeV/c) rather than to a different behavior of low \pt (10-40 GeV/c) vs high \pt ($>$ 100 GeV/c)  jets at RHIC and LHC, respectively.


To summarize most jet properties appear very similar at RHIC and LHC. Differences seen in the fragmentation functions and in the correlation of the track excess at low \pt with respect to the away-side jet cone, need further study to assess their meaning.   

{\centering \section*{6. HEAVY FLAVOR, \jpsi AND $\Upsilon$}}
\label{sec:heavy-flavor}

\begin{figure}[h!]
     \includegraphics[width=75mm]{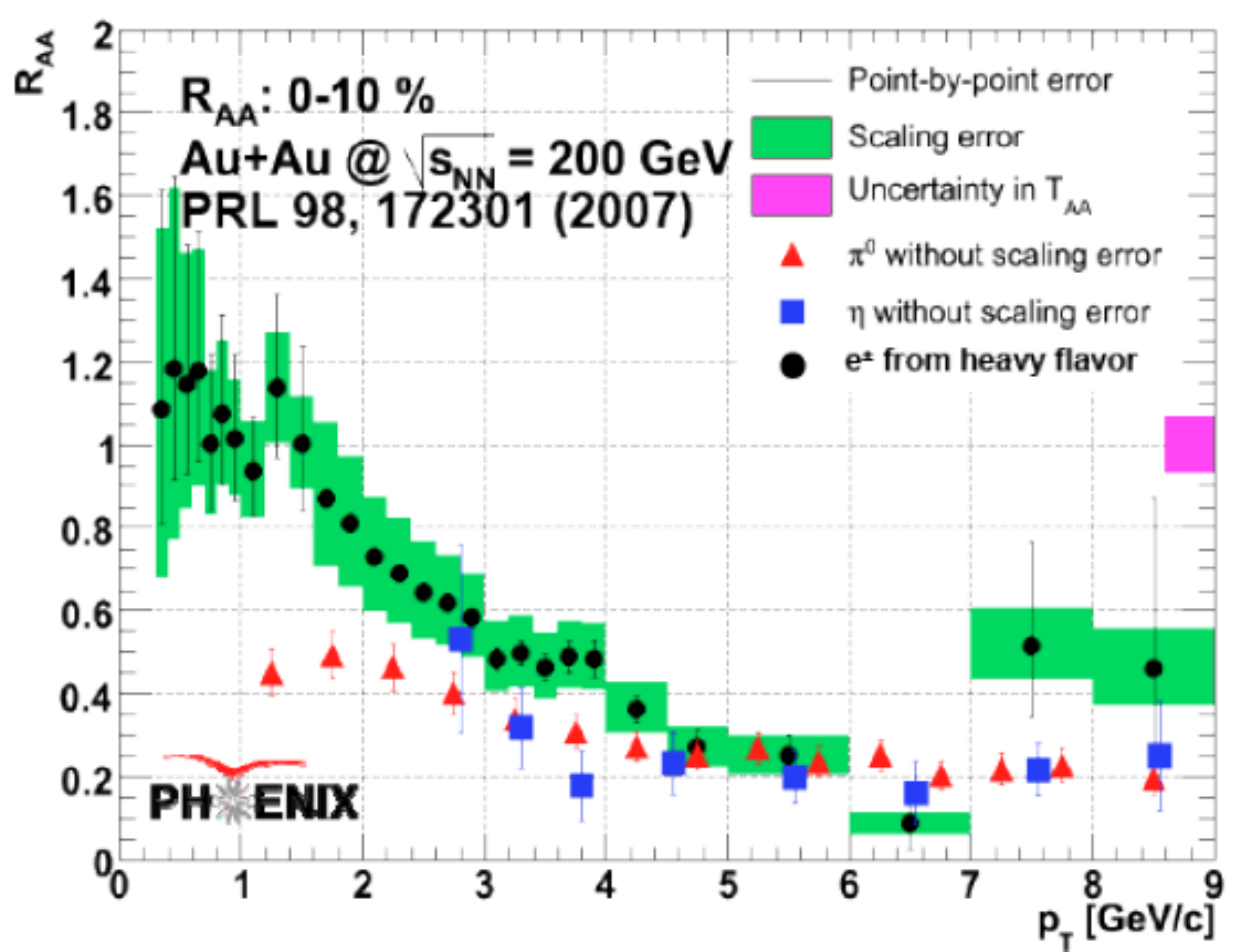}
    \includegraphics[width=75mm]{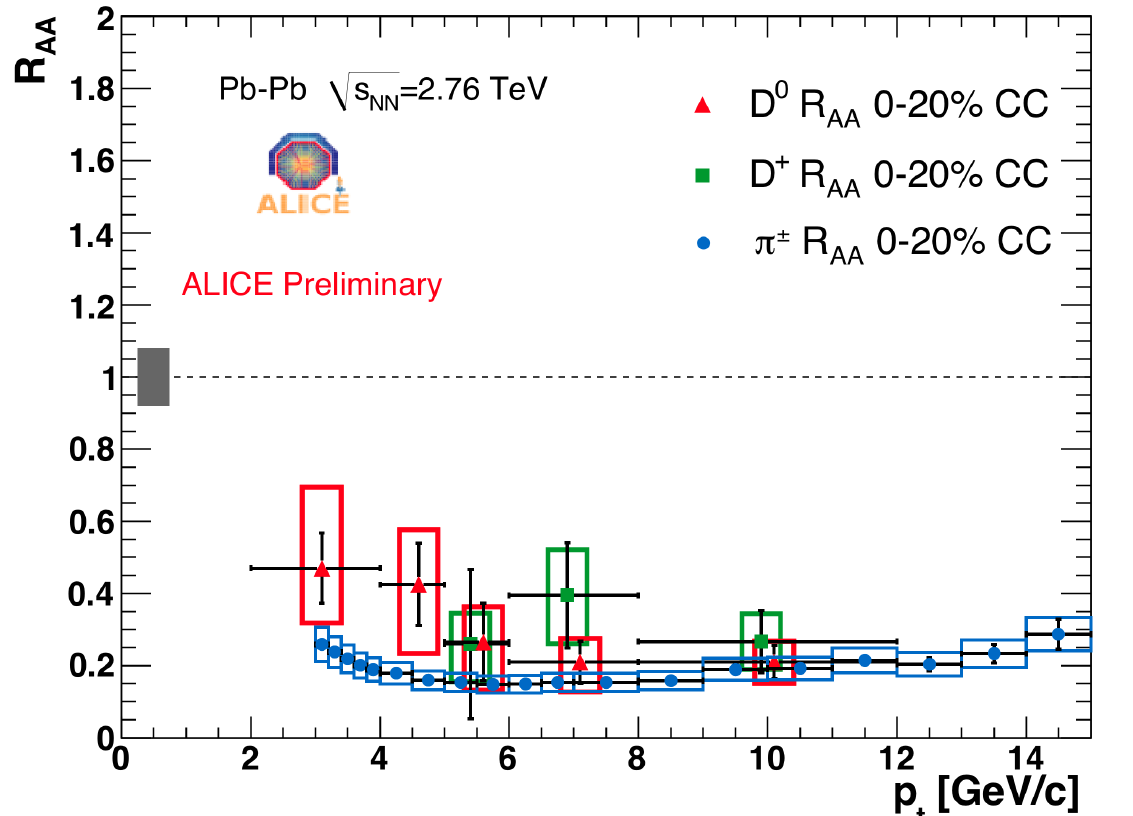}
    \caption{Left: \raa of single electrons from heavy flavor decays in 0-10\% central Au+Au collisions at \sqn = 200 GeV from PHENIX compared with $\pi^0$ results \cite{phenix-raa-hf}. Right: \raa of fully reconstructed D mesons in 0-20\% central Pb+Pb collisions at \sqn = 2.76 TeV from ALICE compared to $\pi^{\pm}$ results \cite{alice-raa-d0}.} 
 \label{fig:charm-raa}
 \end{figure}

$\bullet$  \underline {Open charm.}
One of the most surprising results from RHIC that still remains a challenge for theory, is the large suppression of single electrons from heavy flavor decays. At high \pt (\(>\) 5~GeV/c) the suppression is within errors the same as for light quark mesons (see left panel of Fig.\ref{fig:charm-raa} \cite{phenix-raa-hf}). At lower \pt the suppression is intermediate between that of light mesons and baryons as already discussed in Section 4.

At LHC energies, the large increase of the charm production cross section by more than one order of magnitude allows full reconstruction of charmed mesons via hadronic decays in central Pb+Pb collisions. First results on D$^0$ and D$^+$ \raa from ALICE shown in the right panel of Fig. \ref{fig:charm-raa} \cite{alice-raa-d0}, confirm the large suppression at high \pt (\pt \(>\) 6-7~GeV/c), comparable to the suppression of charged pions, as observed at RHIC. The ALICE data start at \pt = 2 GeV/c and the uncertainties are too large for a qualitative statement on the behavior of open charm \raa at low \pt.

\begin{figure}[h!]
\begin{center}
     \includegraphics[width=72mm, height=55mm]{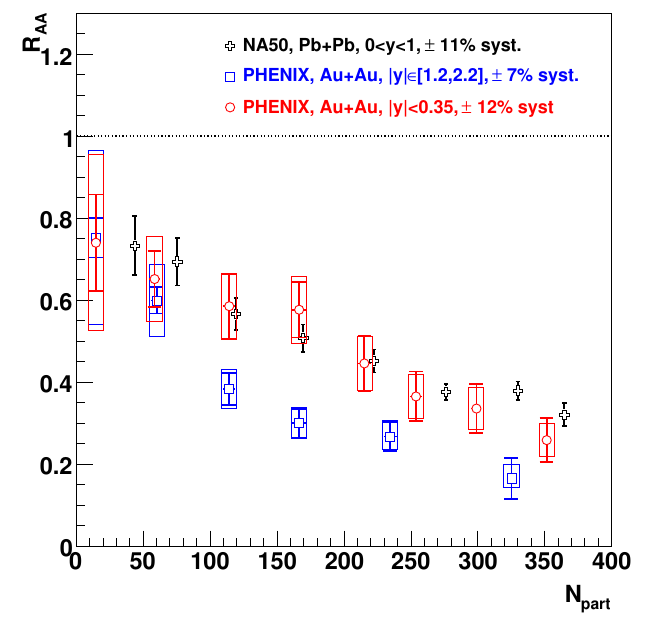}
\end{center}
    \caption{\jpsi suppresion, R$_{AA}$, measured at forward and mid-rapidities by PHENIX in central Au+Au collisions at RHIC compared to SPS results in central Pb+Pb collisions at mid-rapidty \cite{jpsi-phenix-sps}.}
  \label{fig:jpsi-raa-phenix-sps}
 \end{figure}

$\bullet$  \underline {\jpsi suppression.} 
After more than two decades since the classic paper of Matsui and Satz \cite {matsui-satz} that anticipated \jpsi suppression in the QGP due to color screening that prevents \ccbar binding,
the \jpsi saga remains a strong focus of interest. In spite of intense efforts, both experimental and theoretical, a detailed understanding of the \jpsi fate in nuclear collisions is still lacking. The main SPS and RHIC results are briefly reviewed below and confronted with the newly available results from LHC. 
 
Two surprising \jpsi results shown in Fig. \ref{fig:jpsi-raa-phenix-sps} were obtained at RHIC  \cite{jpsi-phenix-sps}. Contrary to simple expectations,  (i) \jpsi is equally suppressed  at RHIC as at SPS and (ii) \jpsi suppression is stronger at forward rapidity than at mid rapidity. 
In addition to the suppression of the \jpsi in the plasma, a variety of other effects have been invoked that are susceptible of affecting the \jpsi yield and could explain the observed results. Among those, one can mention sequential melting of the charmonium states, regeneration of J/$\psi$, gluon saturation, shadowing and cold nuclear matter effects.  A systematic approach involving measurement of the excitation function of \jpsi production in pp, pA and AA collisions will probably be the most effective way to disentangle these various contributions. Data at LHC energies are very valuable in this respect although the first LHC results discussed below appear quite different from those obtained at RHIC.

\begin{figure}[h!]
          \includegraphics[width=72mm, height=55mm]{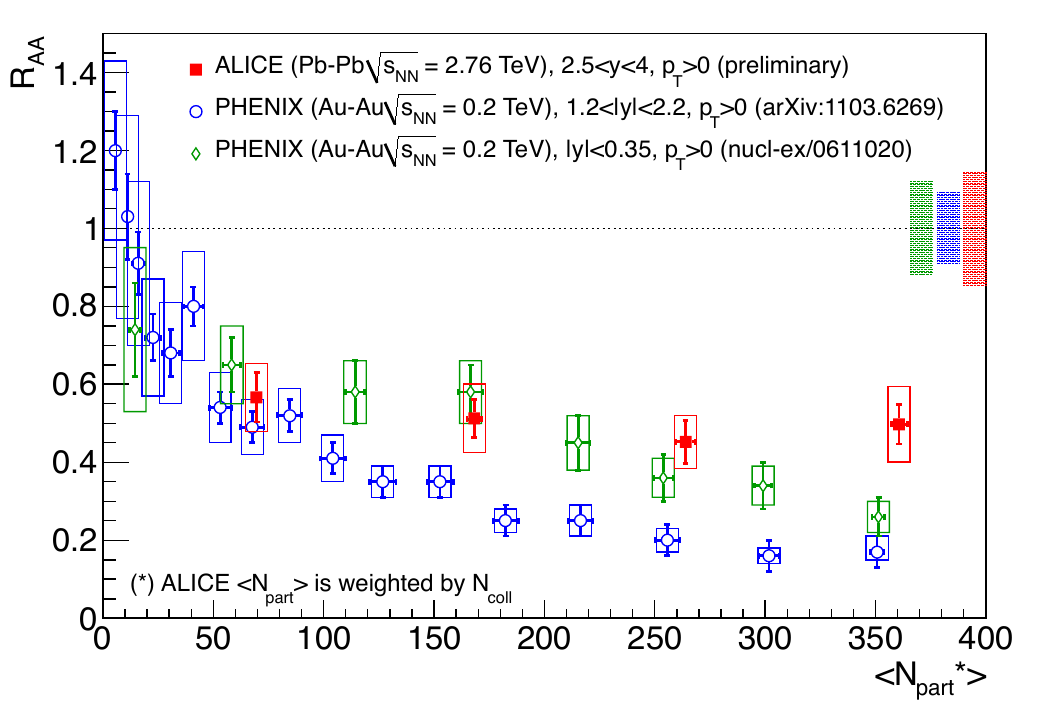}
         \includegraphics[width=72mm, height=55mm]{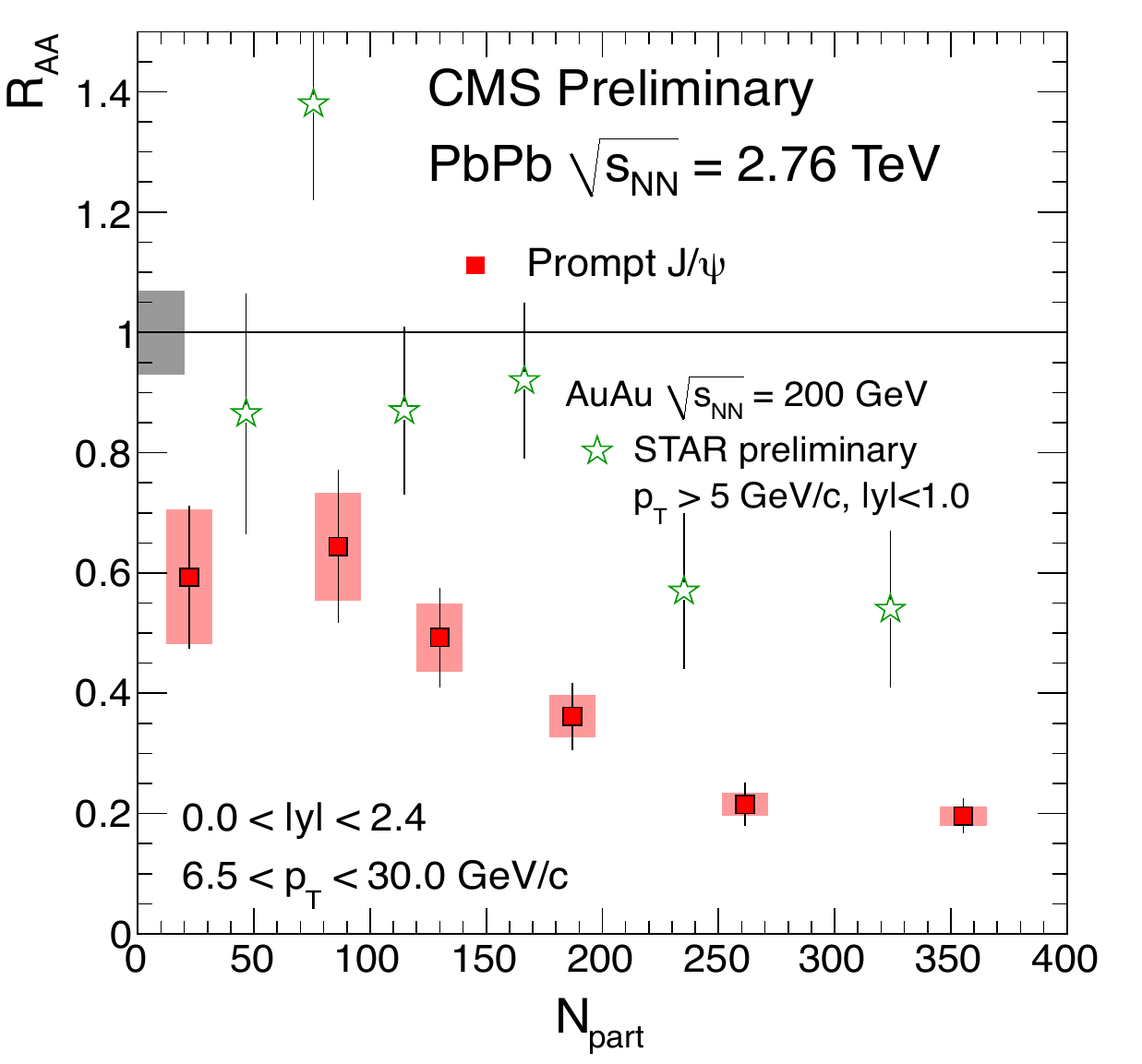}
    \caption{\jpsi \raa vs centrality for \pt\(>\) 0  from ALICE at forward rapidity compared to PHENIX results at forward and mid-rapidities (left panel) \cite{jpsi-raa-alice-phenix} and for \pt \(>\) 6.5 GeV/c at mid-rapidty from CMS compared to STAR results under similar conditions (right panel) \cite{silvestre-qm11}.}
  \label{fig:jpsi-raa}
 \vspace{-6mm}
 \end{figure}

At forward rapidity (2.5 \(<\) y \(<\) 4), the \jpsi \raa for \pt \(>\) 0 measured by ALICE shows very little, or not at all,  dependence on centrality, contrary to RHIC (see left panel of Fig.~\ref{fig:jpsi-raa}) \cite{jpsi-raa-alice-phenix}.  The average magnitude of \raa (in 0-80\% centrality) is $\sim$0.5, a factor of $\sim$2.5 larger than measured by PHENIX  in central collisions at forward rapidity (1.2 \(< |\)y\(| <\) 2.2) and comparable to, or slightly smaller than, the one measured at mid-rapidity (\(|\)y\(| <\) 0.35). 

On the other hand, for high \pt  (\(>\) 6.5 GeV/c), \jpsi \raa results reported by CMS at mid-rapidity  (see right panel of Fig. \ref{fig:jpsi-raa}) \cite{silvestre-qm11} exhibit centrality dependence and a stronger suppression, reaching a value of \raa=~0.2 in central collisions, as opposed to the results at forward rapidity and for \pt\(>\) 0. The suppression is also stronger than the one observed by STAR under similar conditions, displayed in the same figure. 

These first LHC results are intriguing and more Pb+Pb data with higher precision together with reference pp and p+Pb data are needed for further progress on the charmonium states.

\begin{figure}[h!]
     \includegraphics[width=72mm]{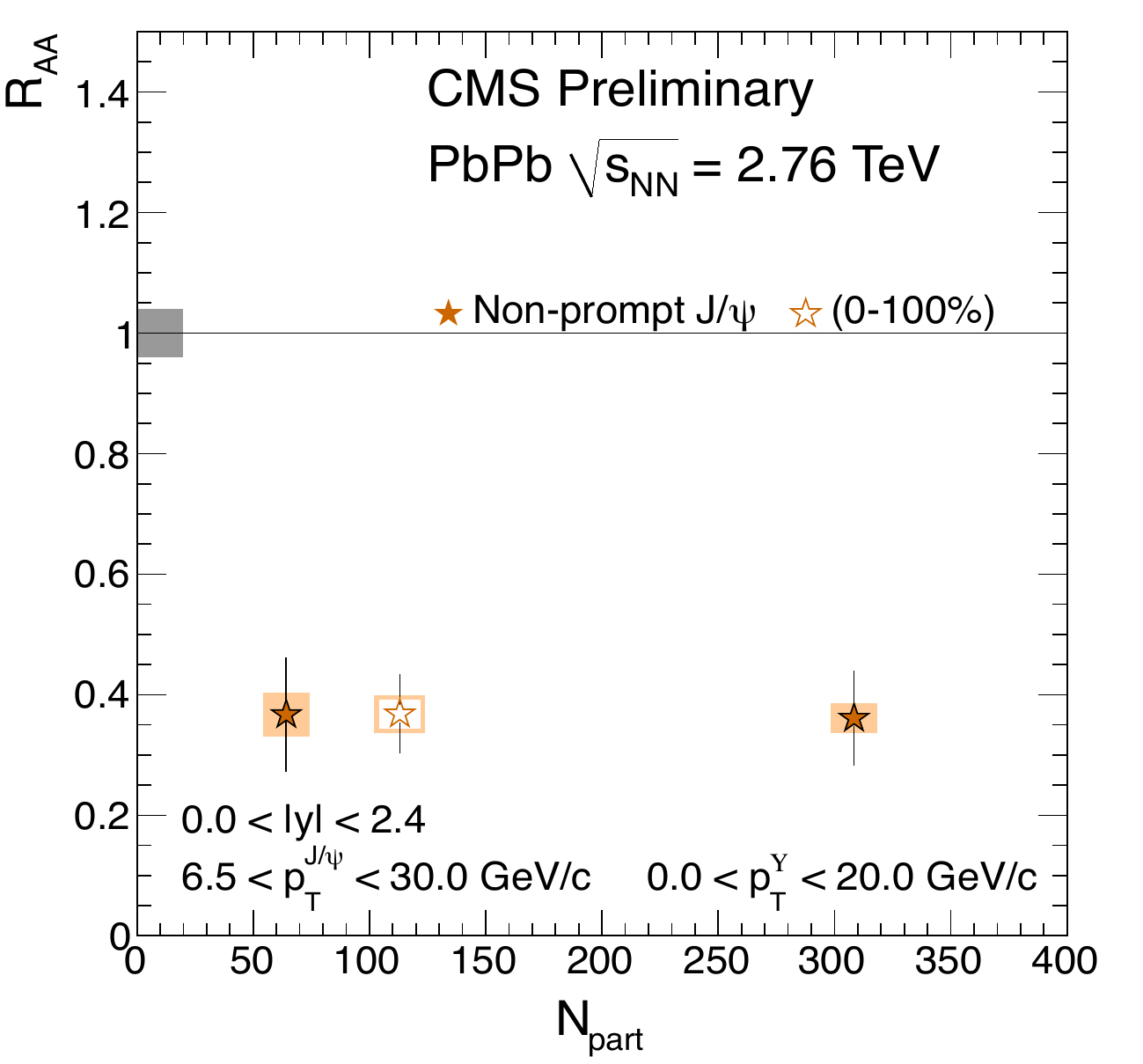}
    \includegraphics[width=72mm]{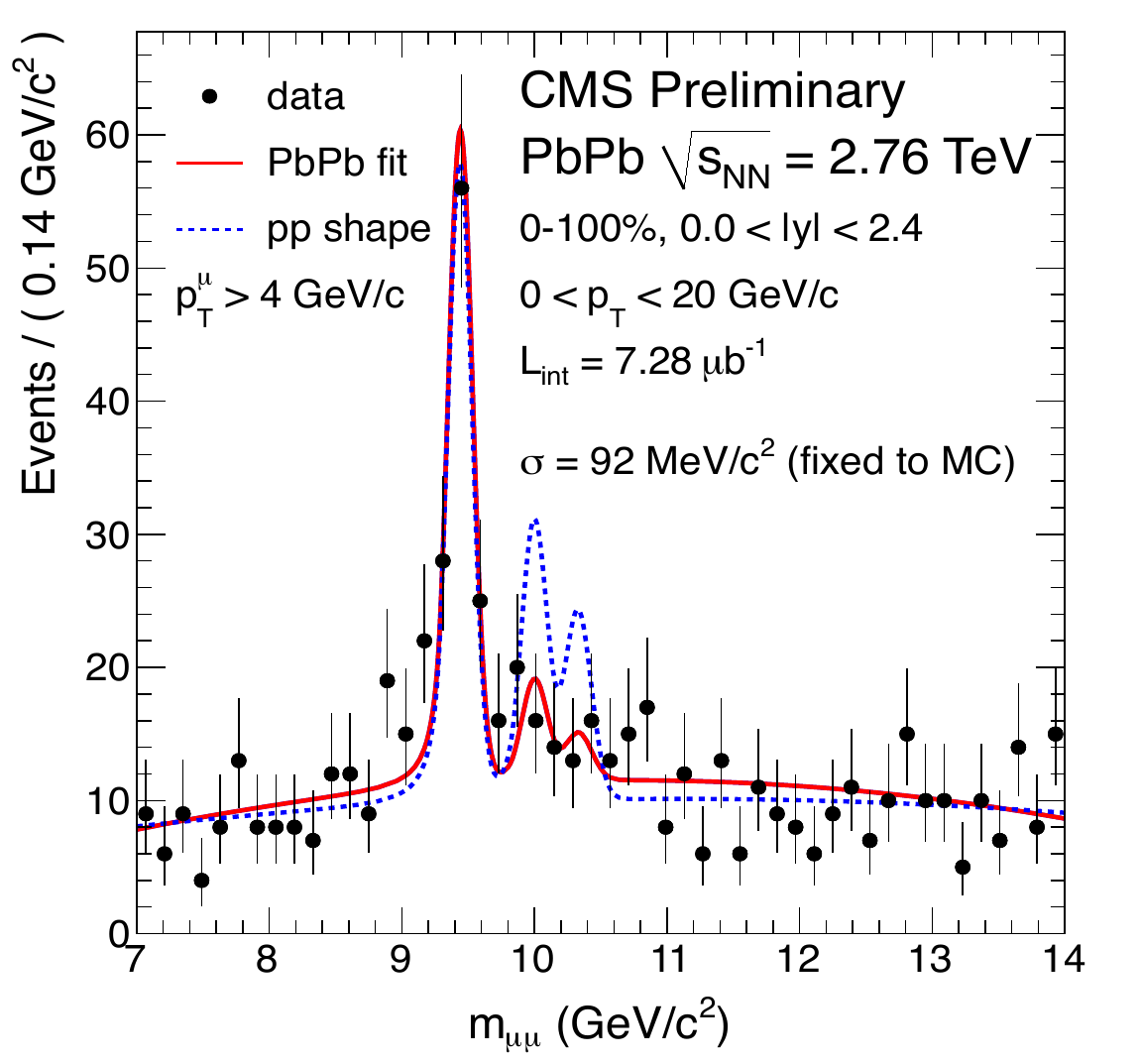}
    \caption{Left: CMS results on \raa of B mesons (non-prompt J$\psi$) \cite{silvestre-qm11}. Right: Dimuon mass spectra in the region of the $\Upsilon$ family measured by CMS in pp and Pb+Pb collisions \cite{cms-upsilon}.}
  \label{fig:bottom-upsilon}
 \end{figure}

$\bullet$  \underline {Open bottom and $\Upsilon$.} 
The high cross sections at LHC allowed already in the first LHC heavy-ion run a glimpse at the suppression of B mesons and the bottomonium family $\Upsilon$ (1S, 2S,3S). B meson production is inferred from the non-prompt \jpsi separated out of the inclusive \jpsi by its displaced vertex. The left panel of Fig. \ref{fig:bottom-upsilon} \cite{silvestre-qm11} shows the non-prompt \jpsi \raa measured by CMS for high \pt (\(>\) 6.5 GeV/c).  A relatively high degree of suppression at $\sim$0.4 is observed suggesting a large energy loss of b quarks in the plasma. No centrality dependence is observed contrary to the prompt \jpsi \raa shown in Fig. \ref{fig:jpsi-raa} (right).

Fig. \ref{fig:bottom-upsilon} (right) shows the \mumu mass spectrum in the region of the $\Upsilon$ family measured by CMS in Pb+Pb (data points and solid histogram) and in pp (dashed histogram) collisions \cite{cms-upsilon}. The two spectra are normalized to the 1S peak. The figure clearly shows that the excited states 2S and 3S are suppressed with respect to the 1S.
 The double ratio [$\Upsilon$(2S+3S)/$\Upsilon$(1S)]$_{PbPb}$/[$\Upsilon$(2S+3S)/$\Upsilon$(1S)]$_{pp}$ gives a value of \(0.31^{+0.19}_{-0.15}(stat.)\pm 0.03(syst.)\).
The $\Upsilon$(1S) state itself is also found to be suppressed to \raa $\sim$0.6. It is interesting to note that a hint of  $\Upsilon$ states suppression with an upper limit of \raa = 0.64 at 90\% confidence level has been reported by PHENIX \cite{phenix-upsilon}.

 With these preliminary results one can foresee a wealth of new and comprehensive studies of the bottomonium states.
 
{\centering \section*{7. SUMMARY}}
The first heavy ion run at the LHC, although of low luminosity, has produced a wealth of high quality results. Most observables are consistent or show a smooth behavior from RHIC to LHC. In addition to considerable quantitative differences due to the higher cross sections at the higher energies of the LHC, there also seem to be some qualitative differences in particular in the behavior of jets and quarkonia. 
 
\vspace{5mm}
{\centering \section*{ACKNOWLEDGMENTS}}
The author acknowledges the conference organizers for the opportunity to present this talk and for a most stimulating conference. This work was supported by  the Israeli Science Foundation and the Nella and Leon Benoziyo Center for High Energy Physics. 
 \vspace{5mm}


\begin{thebibliography}{99}

\bibitem{qm11-proc} See the Proceedings of the 2011 Quark Matter Conference to be published in \JPG.
\vspace{-3mm}
\bibitem{alice-multiplicity} K. Aamodt et al. (ALICE) \PRL {\bf 105}, 252301 (2010).
\vspace{-3mm}
\bibitem{toia-qm11} A. Toia (ALICE) in \cite{qm11-proc}, arXiv:1107.1973.
\vspace{-3mm}
\bibitem{steinberg-qm11} P. Steinberg (ATLAS in \cite{qm11-proc}, arXiv:1107.2182.
\vspace{-3mm}
\bibitem{alice-hbt} K. Aamodt et al., (ALICE), \PLB {\bf 696}, 328 (2011). 
\vspace{-3mm}
\bibitem{alice-flow} K. Aamodt et al., (ALICE)  \PRL {\bf 105}, 252302 (2010).
\vspace{-3mm}
\bibitem{shinichi-qm11} S. Esumi (PHENIX) in \cite{qm11-proc}, arXiv:1110.3223.
\vspace{-3mm}
\bibitem{gong-qm11} X. Gong (PHENIX) in \cite{qm11-proc}.
\vspace{-3mm}
\bibitem{snellings-qm11} R. Snellings (ALICE) in \cite{qm11-proc}, arXiv:1106.6284.
\vspace{-3mm}
\bibitem{alver} B.H. Alver, C. Gambeaud, M. Luzum and J-Y. Ollitrault, \PR {\bf C82}, 034913 (2010).
\vspace{-3mm}
\bibitem{ma} G-L. Ma and X-N. Wang \PRL {\bf 106}, 162301 (2011).
\vspace{-3mm}
\bibitem{jia-qm11} J. Jia (ATLAS) in \cite{qm11-proc}, arXiv:1107.1468 and S. Mohapatra (ATLAS) arXiv:1109.6721.
\vspace{-8mm}
\bibitem{floris-qm11} M. Floris (ALICE) in \cite{qm11-proc}, arXiv:1108.3257.
\vspace{-3mm}
\bibitem{appelshauser-qm11} H. Appelshauser (ALICE), in \cite{qm11-proc}, arXiv:1110.0638.
\vspace{-3mm}
\bibitem{sharma-qm11} D. Sharma (PHENIX) in \cite{qm11-proc}.
\vspace{-3mm}
\bibitem{phenix-jets-raa}  N. Grau (PHENIX) in \cite{qm11-proc}, arXiv:1108.0085.
\vspace{-3mm}
\bibitem{phenix-dijet-correl} J. Jia (PHENIX), \NPA {\bf 855}, 92 (2011).
\vspace{-3mm}
\bibitem{atlas-jets} G. Aad et al., (ATLAS) \PRL {\bf 105}, 252303 (2010).
 \vspace{-3mm}
\bibitem{caines-qm11}  H. Caines (STAR) in \cite{qm11-proc}, arXiv:1106.6247.
\vspace{-3mm}
\bibitem{cms-jet-balance} S. Chatrchyan et al., (CMS) \PRC {\bf 84}, 024906 (2011).
\vspace{-3mm}
\bibitem{bolek-qm11} B. Wyslouch (CMS) in \cite{qm11-proc}, arXiv:1107.2895.
\vspace{-3mm}
\bibitem{phenix-raa-hf} A. Adare et al., (PHENIX) \PRL {\bf 98}, 172301 (2007).
\vspace{-3mm}
\bibitem{alice-raa-d0} A. Dainese (ALICE) in \cite{qm11-proc}, arXiv:1106.4042.
\vspace{-3mm}
\bibitem{matsui-satz} T. Matsui and H. Satz, \PLB {\bf 178}, 416 (1986).
\vspace{-3mm}
\bibitem{jpsi-phenix-sps} R. Granier de Cassagnac, \JPG  {\bf 35}, 104023 (2008).
\vspace{-3mm}
\bibitem{jpsi-raa-alice-phenix} P. Pillot (ALICE) in \cite{qm11-proc}, arXiv:1108.3795.
\vspace{-3mm}
\bibitem{silvestre-qm11} C. Silvestre (CMS) in \cite{qm11-proc}, arXiv:1108.5077.
\vspace{-3mm}
\bibitem{cms-upsilon} S. Chatrchyan et al., (CMS), submitted to \PRL,  arXiv:1105.4894.
\vspace{-3mm}
\bibitem{phenix-upsilon}  E.T. Atomssa, (PHENIX), \NPA {\bf 830} 331c  (2009).

\end{thebibliography}
\end{document}